# Using math in physics:
# 5. Functional dependence

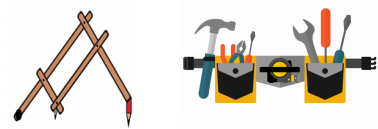

*Edward F. Redish,*
University of Maryland - emeritus, College Park, MD

When students are learning to use math in physics, one of the most important ideas they need to learn is that equations are not just calculational tools; they represent relationships between physical variables that change together (co-vary). How much a change in one variable or parameter is associated with a change in another depends on how they appear in the equation — their *functional dependence*. Understanding this sort of relationship is rarely taught in introductory mathematics classes, and students who have not yet learned to blend conceptual ideas with mathematical symbols may not see the relevance and power of this idea. We need to explicitly teach functional dependence as part of our effort to help students to learn to use math productively in science.

As physics instructors, we typically have a well-integrated knowledge of math and how to use it in science. As a result, we may tacitly assume that functional dependence is easy to read from an equation. We forget how much concept building and blending skills go into the tools we use automatically to analyze these issues.

This paper is part of a sequence on a toolbelt of useful strategies ("epistemic" or "knowledge-building games") that you can add to your physics class to help your students see the value of thinking with math in science and learning how to use math effectively.[1]

Learning to think with math in science is much more than just learning the rules and algorithms of mathematics. One of the most challenging ideas for students is to learn to **blend** their knowledge of the physical world with mathematical symbology.

Functional dependence has many powerful and interesting applications. One of the simplest examples is *scaling*: how something changes when an object's size changes. For example, the volume of a cube ($L^3$) grows faster than its surface area ($6L^2$). The tool icon I have chosen for functional dependence is a pantograph — a pre-computer device for creating scaled diagrams larger or smaller than the original.

## Offering Authenticity

Students who haven't learned to make the physics/math blend may not see the value of thinking with math and resist using symbolic math to reason with. Often, these are students in our service classes who are training for a profession — engineering, health care, or industrial chemistry. Their introductory professional classes often don't use math for anything more than calculation (if it is used at all[2]).

To reach these students and engage them in mathematical thinking, we need to help them see math as *authentic* — of real value to them in their professional lives. One way this can happen is if they see something where their simple non-mathematical intuition is insufficient to help them make sense of what's happening. If math can help them understand a mechanism, to see *why* when previously they only memorized, they often see that as something of authentic value.

A great example is an interview with Ashlyn, a biology student in a study by Watkins and Elby,[3] quoted in the overview paper.[4] At one point in the interview she dismisses mathematical reasoning in the analysis of diffusion, saying "I don't like to think of biology in terms of numbers and variables." She goes on to discuss the specific case of diffusion, where she shrugs off the diffusion spreading equation that tells us that a chemical spreads out diffusively only as the square root of the time: $< (\Delta x)^2 > = 2D\Delta t$. She says

> *if you had a thick membrane and you try to put something through it, the thicker it is, obviously the slower it's gonna go through. But if you want me to think of it as this is x and that's D and then this is t, I can't do it. Like, it's just very unappealing to me.*

But the fact that if you double the distance diffusion needs to travel ($\Delta x$) then the time it takes to get there ($\Delta t$) will be four times as long is of immense importance in biology. The slowness of diffusion is responsible for the evolutionary development of circulatory systems, fractal gas exchange systems (gills and lungs), and electrical neural signalling. (See the sample estimation problem on neurons and *Growing a Worm* in the supplementary materials.)

Ashlyn does in fact appreciate the value of understanding functional dependence mathematically in a different context. Her instructor in a Biological Diversity class had constructed a small wooden model of a horse using dowel legs and a two-by-four wood block body. The legs supported the body with no trouble. But if the linear dimensions of all the parts were doubled, the weight increased by a factor of 8 but the dowel-leg cross sections only by a factor of 4 since the weight grows like the volume but the strength only by the cross-sectional area of the legs. That was not enough strength to hold the weight and the legs broke. Ashlyn was very taken by this, remarking:

> *I remember watching a Bill Nye episode about that, like they built a big model of an ant and it couldn't even stand. But, I mean, visually I knew that it doesn't work when you make little things big, but I never had anyone explain to me that there's a mathematical relationship between that, and that was really helpful to just my general understanding of the world. It was, like, mind-boggling.*

This suggests that if her introduction to the diffusion spreading equation had included its implications for biological structures, she might have appreciated it more. This illustrates the importance of explicitly considering specific examples that your particular population of students see as authentically relevant to their future profession. Functional dependence can be the mathematical tool that enables many of these examples.

## Using functional dependence is not a simple skill

Physics instructors (me too!) often display an equation to a class and make a quick functional dependence argument; for example, "Since Coulomb's law falls off like the square of the distance, these charges have a bigger effect than these." But interpreting that result from looking at an equation requires a number of distinct skills, some of which students in an introductory physics class may have little experience with:

- Parsing what's what
- Understanding dimensional analysis / appreciating quantification

### Parsing what's what

The first task in reading the functional dependence of a quantity from an equation is *parsing* the equation — decoding the role of each symbol. In physics, our equations often have many symbols. Half a dozen variables and parameters are not at all uncommon. Seeing functional dependence and developing an understanding of covariance arising from some relational equation requires not only understanding algebra, it requires an ability to "work the blend" — to identify the various symbols as physical quantities with meaning, not just read them as random symbols. You have to be able to focus on what matters and learn to ignore the rest.

Students often "shut down" when confronted with an equation with many symbols since such equations don't look familiar. If they've seen such complex equations, often it's only been as something to plug numbers into. As a result, students faced with a standard physics equation (that looks simple to a physics instructor) may be unable to bring to bear tools that they can easily use in other situations.

I encountered an example of this in a discussion question I presented to my class in introductory physics for life science (IPLS) students.

> A small spherical moving in a fluid experiences two kinds of resistive forces: an inertial drag force, $F^{drag} = 1/2\ C\rho\pi R^2 v^2$, and a viscous force, $F^{viscous} = 6\pi\mu R^2 v$, where $C$ is the drag coefficient, $\rho$ the density of the fluid, $\mu$ the viscosity of the fluid, $v$ the speed of the object relative to the fluid, and $R$ the radius of the sphere. Is there any speed for which these two forces on the sphere would be equal?

I gave this as a "work together on whiteboards" activity in my IPLS class in the middle of the first semester. As I wandered around watching what students did, I was astonished how many were absolutely frozen and didn't even know how to begin.

I stopped at one student who had not written anything down and asked, "What do you suppose you might do to get started?" She responded tentatively. "I don't know. Look up what some of these numbers are?" What I think was going on here was that because of her unfamiliarity with equations with many symbols (10 in this problem), and her failure to make the physics/math blend, she was unable to identify what were constants and what were variables. She thought that by reducing as much as she could to numbers, the equation might look more familiar.

I decided to help. I asked her to write the equation for each force on her whiteboard. I then said, "These symbols ($1/2\ C\rho\pi R^2$ and $6\pi\mu R^2$) are just constants. Suppose you group them. Call the first set 'A' and the second set 'B'. What do you then have?" She wrote $F^{drag} = Av^2$ and $F^{viscous} = Bv$ and stopped. I asked, "And what do they ask you to do?" She looked, said, "Should I set them equal?" I shrugged. She did and jumped as if shocked. "Oh! It's just $Av^2 = Bv$. That's easy to solve for $v$!" And she proceeded to do so.

The idea that we can choose to temporarily give a combination of symbols a new name and use it to simplify an equation in preparation for "doing math with it" is something many students have never seen. We need to demonstrate it and teach it explicitly. It is an important component skill for picking out functional dependencies.

### Understanding dimensional analysis / appreciating quantification

In addition to simply getting comfortable with equations having many symbols, to interpret functional dependence in an equation students need to learn two deep ideas: (1) that quantities in physics are not numbers but represent physical measurements (dimensional analysis[5]), and (2) that we can build quantitative intuition to get a sense of number and scale (estimation[6]). These two skills are tightly intertwined with an understanding of functional dependence and scaling as can be seen from the examples below and in the supplementary materials. It's important to give students opportunities to build these skills through appropriate tasks and activities in multiple contexts throughout the class.

## Why is functional dependence useful / important?

Three deep insights associated with the way a symbol appears in an equation are helpful in building an understanding of the value of the blend and extracting physical meaning from a complex equation.

- Physical plausibility: Does it make sense?
- Scaling: What happens if you change the size?
- Competing effects: Who's more important?

### Physical plausibility: Does it make sense?

Analyzing the functional dependence of the elements in an equation is one way to help students build the blend of physical intuition and symbolic math. When we look at an equation with many symbols, we can say, "When a quantity x increases and everything else stays constant except quantity y, what happens to y? Does it increase? Decrease? Remain the same? Does that make sense?"

Here's a question that asks students to explicitly generate a physical intuition about the covariance of two quantities and to then check against the equation and see how that information is coded in the equation. This problem combines a question about dimensional analysis with one about physical plausibility. (I give this whether or not we have explicitly discussed capillary action as a topic in the class.)[7]

The strength of the dipole is represented by the combination $p = 2qd$. Some equations giving the force on a 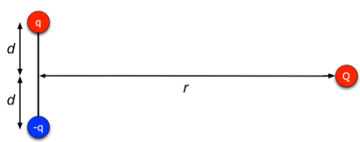 charge $Q$ from a distant dipole are shown below. Some have incorrect dimensions and therefore have to be wrong. Some have implausible functional dependence on a parameter and therefore have to be wrong. (For example: We expect a larger distance $r$ will result in a smaller force. Putting $r$ in the numerator would be a wrong physical dependence.) Put an X in every box that tells us the equation in its row is wrong.

|  | Wrong dimensions | Wrong physical dependence |
|---|---|---|
| $F = k_c p^2 / Q r^4$ |  |  |
| $F = k_c Q p / r^2$ |  |  |
| $F = k_c Q p / r^3$ |  |  |
| $F = k_c Q^3 / pr$ |  |  |

### Scaling: What happens if you change the size?

How a variable or parameter appears in an equation tells how its contribution to the equation changes as it changes its value and therefore how whatever other variable or parameter we are considering must change as well.

Here's a sample two-part homework problem that shows how the idea of scaling can be brought into class with a context that makes it authentic for biology students.

1. A sculptor builds a model for a statue of a terrapin to replace Testudo, the University of Maryland's terrapin mascot. She discovers that to cast her small scale model she needs 2 kg of bronze. When she is done, she finds 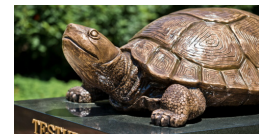 that she can give it two coats of finishing polyurethane varnish using exactly one small can of varnish. The final statue is supposed to be 5 times as large as the model in each dimension. How much bronze will she need? How much varnish should she buy? (Hint: If this seems difficult, you might start by writing a question that is simpler to work on before tackling this one - like a cubical Testudo.)

2. The human brain has 1000 times the surface area of a mouse's brain. The human brain is convoluted, the mouse's not. To see how much of 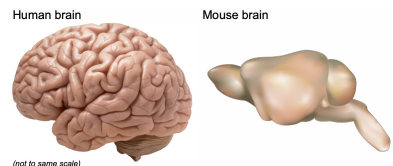 the extra surface is due to convolutions, estimate by what factor the surface area of the human brain would be larger than the mouse's if the human brain were NOT convoluted. How sensitive is your result to your estimations of the approximate dimensions of a human and mouse brain?

This problem combines a real-world case with a life-science authentic one, showing the value of scaling reasoning in both kinds of situations. The first part requires students to understand dimensional analysis: If I'm building a sculpture, the amount of material I need is proportional to the volume, which has dimensionality $L^3$ so this increases with the cube of the sculpture's linear size. But the amount of varnish I need to cover it is proportional to the surface area, which has dimensionality $L^2$, so it only increases with the square of the sculpture's linear size.

The second part of the problem asks students to apply a similar kind of reasoning to a situation that requires an estimation as well as understanding the functional dependence that arises from a dimensional analysis.

Understanding the functional dependence of a system on various parameters is also of particular importance to engineers. Often, large scale systems are too rare, difficult, or expensive to experiment with. What is often done is that experiments are carried out on smaller systems and theoretical functional dependencies are used to scale the results of the measured system up to a real-world size. For engineering students, the scaling aspect of functional dependence is a way of providing examples that engineering students view as authentic.

## Competing effects: Who's more important?

Often, a qualitative analysis of a system will tell us that there are two effects that happen: one that tends to increase the phenomenon we are looking at, the other to decrease it. Without the mathematics, our "fast thinking" response[8] might be to say: "Well, those two effects are likely to cancel." This quick intuition can lead us astray. Looking at the functional dependence (how many powers) of what we are looking at, can help us see which of our two effects is more important. An example of how this works is given in the following problem.

> Smoking tobacco is bad for your circulatory health. The nicotine from tobacco causes arteries to constrict.
>
> The resistance to flow of an artery (symbol Z) follows from the Hagen-Poiseuille equation and is given by $Z = 8\mu L/\pi R^4$ where $\mu$ is the viscosity of the blood, $L$ the length of the artery, and $R$ the radius of the artery.
>
> If the radius decreases by 10%, can you overcome the effect on the resistance by taking a blood thinner to decrease the viscosity by 10%?

The fact that the resistance is inversely proportional to the fourth power of the radius tells us that it will have a much stronger effect on the resistance than a comparable change in the viscosity.

## Functional dependence in class

Students, especially in introductory classes, tend to view what they are learning as independent bits of knowledge. They often fail to generalize from individual examples to general powerful tools. Making the shift from viewing knowledge bits and pieces to developing a coherent framework for reasoning in new situations is perhaps the most valuable shift of perspective that students can learn from a physics class.[9]

Therefore, it's important not only to extract functional dependence from equations throughout the class in multiple contexts, it's important to call your students' attention to the fact that you're doing it. The examples given in the section above each tie functional dependence to other skills and strategies and to authentic implications. It's important to not only give examples like these, but to discuss the "meta-issues" about how we are thinking and what tools we are using to approach these problems.

## Digging deeper: Research resources

The idea of functional dependence emerges from physics education research on identifying covariant dependence[10] and proportional reasoning.[11] The value of authenticity is discussed in case study research with engineering students[12] and with biology students.[13] The concept of "epistemological framing" is discussed in the overview paper to this series[5] and in Hammer et al.[14] Research demonstrating student difficulties with multiple symbols is described in Torigoe & Gladding.[15]

## Instructional resources

Many of the ideas for this series were developed in the context of studying physics learning in a class for life-science majors. A number of problems and activities focusing explicitly on modeling are offered in the supplementary materials to this paper. A more extensive collection of readings and activities from this project on the topic of functional dependence and scaling is available at the *Living Physics Portal*,[16] search "Functional dependence."

## Acknowledgements

I would like to thank the members of the UMd PERG over the last two decades for discussion on these issues. Thanks to Wolfgang Losert for introducing me to the problem of the two resistive forces. The work has been supported in part by a grant from the Howard Hughes Medical Institute and NSF grants 1504366 and 1624478.

---

[1] E. Redish, *Using math in science -- teacher's introduction*, The Living Physics Portal.

[2] For example, most 1000-page introductory biology textbooks for majors do not include a single equation. If they do, those equations are rarely used as part of an explanation.

[3] J. Watkins and A. Elby, Context dependence of students' views about the role of equations in understanding biology, *Cell Biology Education - Life Science Education* 12 (June 3, 2013) 274-286. doi:10.1187/cbe.12-11-0185.

[4] E. Redish, Using math in physics - Overview, preprint.

[5] E. Redish, Using math in physics - 1. Dimensional analysis

[6] E. Redish, Using math in physics - 2. Estimation

[7] Complete solutions for these and the supplementary materials problems are available at the Living Physics Portal.

[8] D. Kahnemann, *Thinking Fast and Slow* (Farrar, Strauss, & Giroux, 2011).

[9] In the research literature, a broad perception as to the nature of knowledge in a given situation is called *epistemological framing*. Helping students shift how they view the knowledge

we are teaching is called *negotiating the epistemological frame*.